\documentclass[a4paper]{article}
\usepackage[a4paper,top=3cm,bottom=2cm,left=3cm,right=2cm]{geometry}

\usepackage[%
backend=biber,
url=false,
style=ieee,
citestyle=numeric-comp,
maxnames=4,
minnames=1,
maxbibnames=99,
giveninits,
uniquename=init,
sorting=none,
dashed=false
]{biblatex}
\bibliography{references}

\usepackage{algorithm2e}
\usepackage{amssymb}
\usepackage{amsmath}
\usepackage{graphicx}
\usepackage{csquotes}

\usepackage{float} 
\usepackage[hidelinks]{hyperref} 
\usepackage{cleveref}
\usepackage[labelformat=simple]{subcaption} 
\crefname{figure}{Figure}{Figure}
\Crefname{figure}{Figure}{Figure}

\graphicspath{{resources/}}

\usepackage{comment}

\newcommand{\vect}[1]{\boldsymbol{#1}}
\newcommand{\papertitle}{Transporting Densities Across Dimensions}

\title{\papertitle}
\author{Michael Plainer\thanks{School of Computation, Information and Technology, Technical University of Munich (\href{mailto:michael.plainer@tum.de}{michael.plainer@tum.de}, \href{mailto:felix.dietrich@tum.de}{felix.dietrich@tum.de}).}
\and Felix Dietrich\footnotemark[1]
\and Ioannis G. Kevrekidis\thanks{Department of Chemical and Biomolecular Engineering and Department of Applied Mathematics and Statistics, Johns Hopkins University and JHMI (\href{mailto:yannisk@jhu.edu}{yannisk@jhu.edu})}}

\begin{document}

\maketitle
\begin{abstract} 
Even the best scientific equipment can only partially observe reality. Recorded data is often lower-dimensional, e.g., two-dimensional pictures of the three-dimensional world. Combining data from multiple experiments then results in a marginal density. This work shows how to transport such lower-dimensional marginal densities into a more informative, higher-dimensional joint space by leveraging time-delayed measurements from an observation process. This can augment the information from scientific equipment to construct a more coherent view. Classical transportation algorithms can be used when the source and target dimensions match. Our approach allows the transport of samples between spaces of different dimensions by exploiting information from the sample collection process. We reconstruct the surface of an implant from partial recordings of bacteria moving on it and construct a joint space for satellites orbiting the Earth by combining one-dimensional, time-delayed altitude measurements.
\end{abstract}

\section{Introduction}
 
Constructing transport maps between distributions is an important problem in many areas of science. The idea is to construct a function that describes an optimal way of mapping one distribution to another.
Applications of such transport maps range from biology~\cite{Moriel.2021, Huizing.2022} over seismic tomography~\cite{Metivier.2016} to finance~\cite{May.2000}. Significant contributions from mathematics include probability and measure theory as well as optimal transport theory~\cite{Villani.2003, Villani.2009}.
The first solution to the optimal transport problem was proposed by Monge, who suggested individually mapping points of the input distribution to points of the output distribution via a transport map that minimizes the distance in the input and target space~\cite{Monge.1781}. However, this problem definition is ill-posed; a description of the problem based on work of Kantorovich then resolved this issue~\cite{Kantorovich.1942, Kantorovich.1948}. The main idea is to minimize over joint measures between source and target distributions instead of maps.
Nowadays, depending on the concrete cost function of the transport, many concrete algorithms to solve specific transport problems have been developed~\cite{Altschuler.2017, Mrigot.2021}.

\begin{figure}[h!]
    \centering
    \includegraphics[width=1\textwidth]{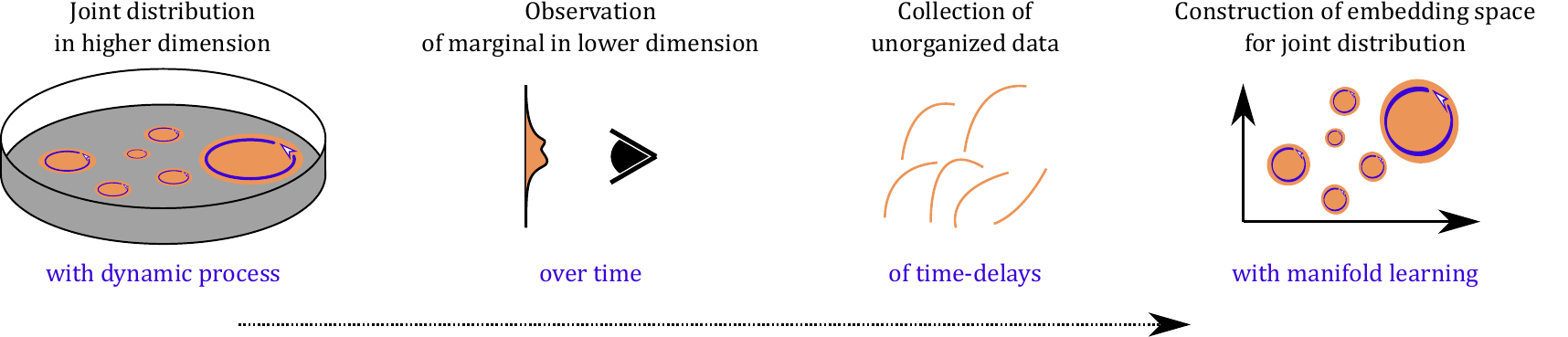}
    \caption{Illustration of our approach to transport from a marginalized distribution in a low-dimensional space to a joint distribution in a higher-dimensional space. A higher-dimensional joint density explains the underlying process (in this instance, bacteria moving in a petri dish), but only lower-dimensional marginal densities can be observed. When time-series of the measurements from the data-collection process are available, instead of just static measurements, our method can reconstruct a joint density on a space diffeomorphic to the original.}
    \label{fig-1:illustration}
\end{figure}
In our work, we follow this common view and treat both source and target distributions as marginals of a joint distribution, which is to be constructed (compare \cref{fig-1:illustration}).
However, different from the Kantorovich formulation of optimal transport, we do not freely optimize over all joint distributions to minimize an objective. Instead, we assume that both source and target distributions are marginals of a shared unknown but \emph{fixed}, higher-dimensional joint distribution that is to be discovered by our procedure. Importantly, as will be demonstrated, the commonly used Kantorovich solution (i.e., optimal transport) may not agree with the given, fixed, joint distribution, even though the marginals (source and target) of both solutions are identical. 
Determining the correct mapping is especially important in areas such as domain adaptation, where it is crucial that the correct solution is found~\cite{Courty.2014}.

What allows us to uncover a unique joint density is that we require access to {\em additional information} about measured points in any of the given marginal distributions: we assume that the data in the marginals were collected by running a time-evolving  data collection process (an {\em observation process}) on the joint distribution (e.g., on a joint object), and the points on the marginal (e.g., on one of the coordinate axes) were recorded (and are accessible) as a short time series.
For example, if there is a dynamic process with a stationary distribution on the joint object, we would not only keep individual measurements to create the marginal distributions. Instead, we must keep {\em short sequences of the time-delayed (marginalized) measurements}. This approach is motivated by Whitney's and Takens' embedding theorems~\cite{Whitney.1936, Takens.1981}. 
We show that if the time-delayed information is kept instead, and processed further using manifold learning techniques, we can construct a one-to-one copy of the joint distribution.
With our approach to approximate the fixed underlying joint distribution, we can find a transport even for source and target distributions defined on spaces of different dimensions. This is contrary to typical density transport approaches~\cite{Tabak.2010, Tabak.2013, Peyre.2019}, which are only applicable when the source and target dimension match.

In previous work~\cite{Moosmuller.2020}, time-delay embeddings were used to transport distributions with singular and discontinuous densities.
Here, we build on this work and focus on transporting distributions across spaces of different dimensions, namely the marginals of a common fixed but a priori unknown distribution in a higher-dimensional space.
While this may sound contrived, we will demonstrate and explain the procedures in different scenarios.

Our main contributions in this work are as follows.
In section~\ref{sec:transport across dimensions framework}, we define a specific mapping from measurements (that typically lead to marginal densities) back to an underlying joint density.
We demonstrate the process by reconstructing a two-dimensional surface by observing the motion of bacteria on a 3D implant (section~\ref{sec:bacteria on surface}). Our approach allows us to reconstruct a diffeomorphic (i.e., similar) copy of the original, two-dimensional implant surface even in a challenging example where we only have a one-dimensional measurement for each bacterium at a specific time (namely, we can only observe a single coordinate).
We then show how this procedure can be extended to data from multiple, possibly noisy sensors to find an underlying, shared space in section~\ref{subsec:jointly-smooth-functions}. For this, we utilize jointly smooth functions~\cite{Dietrich.2022}, which allow us to determine common features in multimodal data.
Directly constructing those functions, as discussed in~\cite{Dietrich.2022}, does not yield the true underlying common features when they are transformed by non-invertible functions. A better common representation can be found by first embedding the features into a higher dimension (using their time-delayed measurements) and then transporting. We demonstrate the pipeline on altitude measurements from two satellites and show that they are sufficient to reconstruct a joint distribution embedded in three-dimensional space. Using this distribution, we can reconstruct the densities of both satellites through marginalization.

\section{Related work}
While research has been conducted on (optimal) transport with mismatching dimensions based on choosing a cost~\cite{McCann.2020, Chiappori.2017, Gangbo.2000, Lott.2016}, it is still an open problem to find transport maps across different finite dimensions~\cite{Chiappori.2017}. Some authors even believe that the regularity of transport maps between mismatching dimensions is the biggest issue in the field in general~\cite{Chiappori.2017}. As such, much of the work done in this direction focuses on the underlying properties of those maps~\cite{McCann.2020, Gangbo.2000} instead of a procedure to construct them. Most notably, \cite{McCann.2020} show that an optimal transport between unequal dimensions is equal to a solution of a particular form of the Monge-Ampère equation~\cite{Philippis.2014}. In a sense, this gives a way to solve the transport, at least in simple scenarios.

When the marginal variables are independent of each other, the reconstruction of the joint distribution becomes trivial, as it is just a multiplication of the marginal densities. When the observed marginal variables are \emph{categorical}, they can be combined into a joint distribution by simple mathematical operations~\cite{CastellanJr.1970439}. On the other hand, methods constructing joint distributions between clean and corrupted data sets often assume that the data domains are conditionally independent to simplify the process~\cite{Zheng2022}.

Even when the variables are not independent, the joint probability density function of multidimensional variables can be decomposed into the continuous product of a marginal density function and additional, bivariate copula functions through ``Vine copula''. 
Based on this, multidimensional dependent problems can be transformed into two-dimensional dependent problems, reducing the dimensionality. An approach using this idea for sensitivity analysis is presented in~\cite{Bai2021}.
The ideas of couplings can also be used to estimate the joint probability density using parameterized models and maximum likelihood estimation, as was done in~\cite{Mu20201}.

Often, analytically tractable formulas for the joint probability density do not even exist. An analytical solution for linear constraints between marginals is already infeasible. Instead, Markov chain Monte Carlo (MCMC) methods can approximate the joint probability density from marginals~\cite{Cencic2015196}. Those methods can also be used to estimate the posterior density function based on samples from the joint posterior distribution~\cite{Oh1999411}.
Other models for the joint distribution function include the so-called Nataf models, which are distribution models of specific joint distributions consistent with prescribed marginal distributions and a correlation structure~\cite{Choi20081171}.
For several known marginal probability distributions, the joint distribution can be constructed with the sum-of-uniforms method~\cite{Chen2005226}.
More recently, support vector machines have been used to parameterize more complicated joint distribution functions with high accuracy~\cite{Shan2011140}.

However, there are also different approaches that rely on measurements of the same instances over a period of time. 
For example, learning transformations to a latent distribution has been used to facilitate predictions of time-dependent distributions~\cite{Lu2022}.
In clinical studies, the data is typically incomplete or censored (e.g., patients drop out of studies), but the same variables are repeatedly measured over a longer frame of time (e.g., CD4 cell count in AIDS patients). In such cases, the joint probability can also be estimated~\cite{Hogan1997239}.

\section{Methods}\label{sec:transport across dimensions framework}

\subsection{Marginal and joint densities}\label{sec:def marginal and joint}
Densities play a crucial part in probability theory and optimal transport~\cite{Villani.2009}. For discrete random variables, the probability mass function \(p(x)\) describes the probability for a given event \(x\) to occur. Probability density functions (PDFs) are the continuation of this idea to the continuous case. Unfortunately, densities do not have such a nice and intuitive meaning as probability mass functions. For example, the probability density function \(p(x)\) can be larger than \(1\), which of course, cannot describe the probability of an event. Still, a higher density means that an event is more likely to occur compared to an event with a lower density. For a probability density function $p$ to be valid, it must hold that
\begin{equation}
	p(x) \geq 0 \quad \text{and} \quad \int_{-\infty}^{\infty} p(x)\, d\lambda(x) = 1 \text{,}
\end{equation}
where \(\lambda\) is the Lebesgue-measure.

Density functions can also be used to describe the probability of multiple random variables to occur concurrently. \(p(x,y)\) for example describes the density for the event \(x\) and \(y\) and is thus called the \emph{joint} probability density function of \(x\) and \(y\). For example, it could describe the outcome of rolling two dice. Such a joint PDF can again be used to determine the density of a single random variable \(x\) (e.g., a single die) by integrating over all possibilities of \(y\) such that 
\begin{equation}
	p(x) = \int_y p(x,y)\, d\lambda(y) \text{.}
\end{equation}
As \(p(x)\) results from a joint density and only depends on a subset of the variables, it is called \emph{marginal} density. While joint PDFs can describe the random process and all included variables completely, in practice, they are often simply unavailable or not tractable to observe or compute. Processes, where only a subset of (random) variables can be recorded, exhibit an intuitive scenario where marginal densities arise. The aim of this paper is to demonstrate how the joint PDF can be reconstructed when only observations/measurements of the marginal density are available. The concept of marginalized and joint densities can also be extended to higher-dimensional densities, such that
\begin{equation}
	p(x,z) = \int_u \int_y p(u,x,y,z)\, d\lambda(u)\, d\lambda(y) \text{.}
\end{equation}

\subsection{Transporting densities}\label{sec:transport densities}
The field of transportation theory studies how one density (or distribution) can be transformed into another. With Monge's formulation of this problem~\cite{Monge.1781}, this procedure relies on a transport map \(T\) that moves individual mass from one distribution \(\mu\) to \(\nu\) while still ensuring that the properties of a distribution are retained. This mapping \(T\) specifies, for each sample \(x \sim \mu\), where it should be transported to, so that \(T\left( x \right) = y \sim \nu\). In other words, the function \(T\) defines, for each point, where it should be in the other probability space so that the result follows the target distribution \(\nu\).
More concretely, Monge's formulation of the transportation problem, dating back to 1781, defines such a transport. At that time, the goal was to find an optimal way to fill a hole in the ground with shovels from a heap of soil~\cite{Monge.1781, Villani.2003, Villani.2009}. In this analogy, the \(x\) and \(y\) would be the position on the heap and the hole, respectively. A more up-to-date example would be to find a map \(T\) that transports the points of a uniform density \(\mu\) to a normal density \(\nu\). 

As there are many possible solutions, usually an additional optimality constraint is added where a Euclidean distance is minimized (i.e., minimize the walking distance with shovels full of dirt).
To formalize this concept, \(\nu\) is the push-forward measure of \(\mu\) under \(T\) if
\begin{equation}
	\label{eqn:push-forward}
	\forall A \quad \nu \left( A \right) =  \mu \left( \left\{ T\left( x \right) \in A \enspace \middle| \enspace x \in X \right\} \right) = \mu \left( T^{-1} \left( A \right) \right) \text{.}
\end{equation} 
While this formulation might seem daunting, it simply describes that every volume of soil in the filled hole must come from an equally sized volume from the heap. In other words, this notion enforces that the map \(T\) is \emph{measure preserving}.

As this work focuses more on densities than measures, we reformulate equation~\ref{eqn:push-forward} by the density of distribution $\nu$ with respect to a reference distribution $\lambda$, to
\begin{equation}
	\nu \left( A \right) = \int_A p_\nu \left( y \right)\, d\lambda \left( y \right) = \int_{T^{-1}(A)}p_\mu \left( x \right)\, d\lambda \left( x \right) = \mu \left( T^{-1} \left( A \right) \right) \text{.}
\end{equation}
The densities $p_\nu$ and $p_\mu$ are thus also related by the map $T$.

\subsection{Transporting marginal to joint densities}\label{sec:transport marginal to joint}
Previous work~\cite{Moosmuller.2020} discussed how {\em an observation process} can be employed to recover the underlying manifold. The focus of that publication was on the transport of singular and/or discontinuous densities, which can result from non-invertible {\em observation functions}. In this work, we use the same idea to construct a meaningful transport of densities on spaces of different dimensions, including transporting marginal to joint densities.

The ideas in~\cite{Moosmuller.2020} rely on data where not only a single observation is available but a short sequence of measurements. For example, instead of taking a single picture of static objects, one could let a camera move with constant speed, and take multiple pictures in a changing environment. This gives rise to so-called time-delayed observations. Equivalently, this can be seen as objects moving on a manifold, all moving on non-intersecting trajectories. Note that this observational process does not necessarily require movement, but could rely on a change in any variable. If this variable is time, the measurement changes with time, and gives rise to time-delayed measurements of the process of collecting a data point. For our method, ordered series of measurements \(( y^{(i)}_1, y^{(i)}_2, \dots\ ) \) instead of individual measurements \(y^{(i)}\) are required; it is this additional information that enables our entire workflow. 

With sufficiently many sequential measurements, a diffeomorphic manifold to the one underlying the observations can be constructed~\cite{Moosmuller.2020}. Takens' theorem~\cite{Takens.1981} specifies that the underlying manifold can be embedded with at most \(2d + 1\) observations. Thus, when the original data was sampled from objects moving on a 2D manifold, $5$ time-delayed measurements \( (y^{(i)}_1, \dots, y^{(i)}_5) \) are sufficient to provide a diffeomorphic embedding. We proceed to embed the diffeomorphic manifold in this higher-dimensional space.

When the intrinsic dimension of the lower-dimensional manifold embedded in a higher dimension is known, we can try to parameterize the higher-dimensional embedding with the correct number of parameters. For this, any (non-)linear dimensionality reduction mechanism can be used, such as Diffusion Maps (DMAPs)~\cite{Coifman.2006}. When done properly, this reconstructs the true manifold (up to diffeomorphism) and in a sense \emph{unfolds} the manifold. For some physical processes, the intrinsic dimensions might be known, in other cases, one might be able to determine it visually (especially in lower dimensions). The more general and robust solution is to rely on dimension estimation algorithms such as~\cite{Rozza.2011, Campadelli.2015}.

Once the joint distribution is available, the points on the joint base space can be transported with any classical transportation algorithm. This is possible because our procedure increased the dimension and thus implicitly allows transport from marginal to joint densities (i.e., from a low to a high dimension). In this work, we typically leave out the transports that could be done after the transformation and will focus on {\em transporting marginal to joint densities}. In the instances where we do an additional transport, we rely on Normalizing Flows parameterized by neural networks, which are described in appendix~\ref{apx:normalizing-flows}.

\section{Results}

\subsection{Same dimensional transport}
\label{subsec:1d-transport}
We will begin demonstrating the procedure with a one-dimensional example similarly to~\cite{Moosmuller.2020}, but we will discuss an actual parametrization of the map \(T : x \mapsto y\) that we will reconstruct with a Normalizing Flow. While this basic example does not perform a transport to a higher dimension, it explains the fundamentals of the underlying procedure. In the first part, the general setting will be introduced, and since the dimensions match, we will compare it to applying a classical optimal transport algorithm directly. However, a classical approach will typically not be able to recover the underlying transport map.

Assume that samples \(y^{(i)} \in \mathbb{R}\) are observed. While the underlying distribution of those samples themselves is not known analytically, it is known that data was produced by sampling \(x^{(i)} \in \mathbb{R}\) from a uniform distribution and the samples \(x^{(i)}\) were transported by some function \(T: x \mapsto y\). The goal is to find the underlying transport function \(T\) and \(T^{-1}\), mapping between the spaces. While this may appear rather abstract, a simple use case could be based on uniformly distributed cars driving on a highway. Each car has a thermometer with \(y^{(i)}\) being the recorded temperature of the car \(i\). The cars lack GPS, and thus, the \(x\)-position on the road is unknown.

\subsubsection{Construction of the classical transport solution}
\begin{figure}[htbp]
    \centering
    
    \begin{subfigure}[t]{0.33\textwidth}
        \includegraphics{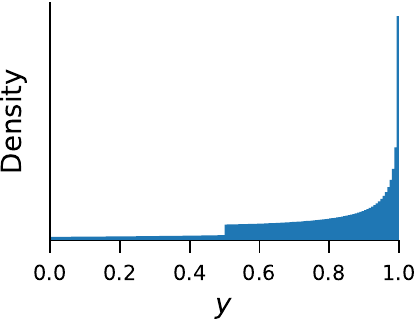}
        \caption{Histogram of observations}
        \label{fig:1d-transport-map-comparison-histogram}
    \end{subfigure}
    \begin{subfigure}[t]{0.33\textwidth}
        \includegraphics{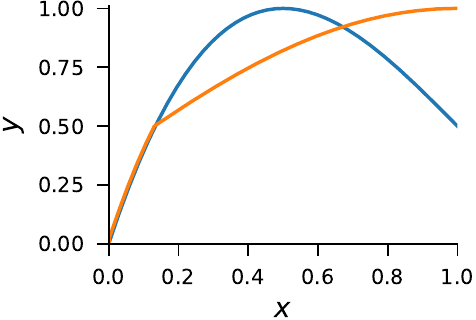}
        \caption{Different transports to \(y\)}
        \label{fig:1d-transport-map-comparison-maps}
    \end{subfigure}

    \caption{Density reconstruction in one dimension. In \subref{fig:1d-transport-map-comparison-histogram}, the density of the samples \(y^{(i)}\) that were observed is plotted in the form of a histogram. In \subref{fig:1d-transport-map-comparison-maps}, two different solutions are illustrated that map samples from \(x\) to \(y\). The orange function is invertible and can be found by a Normalizing Flow or by solving the optimal transport problem, while the blue function is smooth and is the true \enquote{unknown} \(y = T(x)\). Both explain the observed density when assuming a uniform base density.}
    \label{fig:1d-transport-map-comparison}
\end{figure}

We will assume a concrete instantiation of the problem with the \enquote{unknown} function \(T\left( x \right) = -2(1-x)^3 + 1.5 (1-x) + 0.5\) and the underlying space \(x\) being uniform on \([0, 1]\). Any successful transport from a uniform density to the observed density yields such a mapping \(T\). We will rely on Normalizing Flows to transport a uniform density to the observed density given by \(y^{(i)}\). This has been visualized in \cref{fig:1d-transport-map-comparison-maps}~(orange curve). Since Normalizing Flows are inherently invertible, they can only recover invertible functions. This solution is also found when solving the classical optimal transport problem~\cite{Moosmuller.2020}.

In many cases, finding any transport map \(T\) is sufficient, for example, when the goal is to draw new samples from the observed distributions. In the analogy with the cars measuring the temperature on the highway, the goal is to find the true underlying map. The two maps \(T_1, T_2\) predict different temperatures for most positions on the highway, and most likely, only the true map is useful.

This simple example illustrates a challenge 
with Normalizing Flows: They cannot be used directly to find non-invertible transport functions. Similar problems arise with other classical transport solutions~\cite{Moosmuller.2020}. The second and more severe limitation of classical approaches is that the source and the target dimension need to match. Expanding the previous example from cars on a highway to ships sailing a two-dimensional sea where only the temperature can be recorded, a Normalizing Flow is unable to find \emph{any} solution. This is because the temperature is one-dimensional, but the underlying space is two-dimensional.

\subsubsection{Transport with time-delayed observations}
\begin{figure}[htbp]
	\centering
    \begin{subfigure}[t]{0.3\textwidth}
        \includegraphics{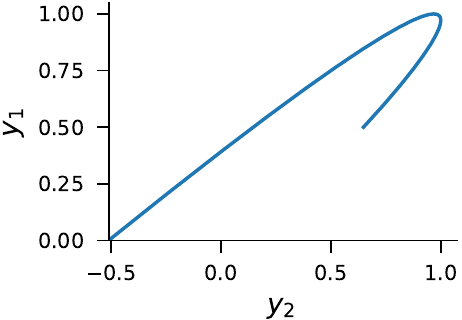}
        \caption{Embedding of time-delayed observations}
        \label{fig:1d-unfolding-embedding}
    \end{subfigure}
    \begin{subfigure}[t]{0.3\textwidth}
        \includegraphics{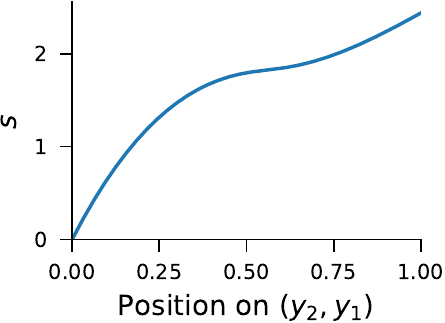}
        \caption{\label{fig:1d-transported}Arc length of \(y_2, y_1\)}
        \label{fig:1d-unfolding-alen}
    \end{subfigure}
    \begin{subfigure}[t]{0.3\textwidth}
        \includegraphics{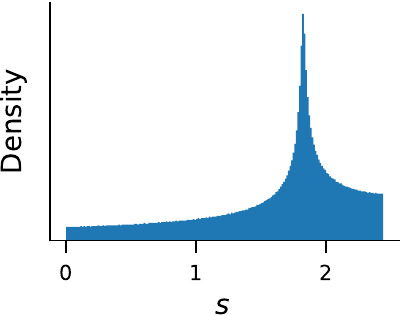}
        \caption{Density of arc length}
        \label{fig:1d-unfolding-alen-density}
    \end{subfigure}
    
    \caption{Unfolding of a non-invertible transport map. \subref{fig:1d-unfolding-embedding} shows the time-delayed observations revealing a diffeomorphic version of the hook. In \subref{fig:1d-unfolding-alen}, the arc length of the diffeomorphic embedding is visualized, which is an ``unfolded'' (i.e., invertible) version of the hook. \subref{fig:1d-unfolding-alen-density} shows the density of the arc length $s$, which forms the basis for the transport.}
    \label{fig:1d-unfolding}
\end{figure}

We will now assume access to {\em an observational process}, meaning that time-delayed measurements are available such that \((y^{(i)}_1, y^{(i)}_2) = (T(x), T(x + \tau))\) for an arbitrary but fixed \(\tau>0\). Continuing the analogy from before, the cars still do not have any GPS, but now they all move at the same speed on the highway and record the temperature twice. Shortly after the first measurement \(y^{(i)}_1\) has been taken, the temperature is recorded again, resulting in \(y^{(i)}_2\). Plotting these observations against each other reveals a function that is very similar (i.e., diffeomorphic) to the real manifold. While the function is a bit shifted, the key defining feature, the hook, is already present. \Cref{fig:1d-unfolding}(a) visualizes this.

Previously, we were able to learn a Normalizing Flow to transport samples from \(y \in \mathbb{R}\) to \(x \in \mathbb{R}\) because the dimensions match.
Now, the dimension of the data has been artificially increased to two dimensions by collecting temporal data, from \(y^{(i)}\) to \((y^{(i)}_2, y^{(i)}_1)\). With this mismatch in dimensions, such a transport is not possible anymore. To conclude the procedure, we must find a way to parameterize the one-dimensional manifold (i.e., the line) embedded in two dimensions by a single parameter. 

The arc length can be used to describe one-dimensional manifolds with a single parameter. For later higher-dimensional examples, it will be more difficult to find a correct-dimensional embedding. The arc length of a function \(c\) is defined as \(\text{arcl}\left( x \right) = \int_{0}^{x} \| c'(t) \| dt\) and can be imagined as measuring the length of the function while drawing it. With this intuition, it becomes clear that we can use this position on the line as an {\em intrinsic} single parameter to identify a point (i.e., there exists a diffeomorphic mapping). For example, the information \enquote{at $40$\% of the line} gives an exact position. The arc length \(s\) applied to the diffeomorphic embedding of the hook can be seen in \cref{fig:1d-unfolding-alen}.

With this procedure, the dimension of the lower-dimensional representation of the diffeomorphic embedding (i.e., the arc length \(s\)) and the dimension of the original \(x\) space again match. This allows us to apply a Normalizing Flow to learn the mapping from the density given by the arc length representation \(s\) to a uniform density \(x\). The resulting density from the arc length is represented in \cref{fig:1d-unfolding-alen-density}, which can be transported with a Normalizing Flow to a uniform density on \(\left[ 0, 1\right]\).

Using a Normalizing Flow instead of a classical transport algorithm gives access to the map \(T\) and its inverse \(T^{-1}\). To reconstruct the map \(T\), we can apply the inverse of the Normalizing Flow to samples \(x\), giving us a respective arc length value \(s\). This, on the other hand, corresponds uniquely to a position on the delay embedding \((y^{(i)}_1, y^{(i)}_2)\), yielding a mapping from \(x \to y\) which aligns with the true underlying map in \cref{fig:1d-transport-map-comparison-maps} (blue curve). In the car example, this would conclude the reconstruction of a function that maps the position of the highway to a concrete temperature. For the inverse, mapping \((y^{(i)}_1, y^{(i)}_2)\) to \(x\), we calculate the arc length up to the observations and apply the forward transform of the Normalizing Flow. Although normalizing flows are only capable of finding invertible solutions, with this reparametrization we were able to find a mapping that is \enquote{non-invertible} in the original space, but invertible with the increase in dimension. 

Note that without any additional information about the true underlying space, our approach can only be used to reconstruct it up to diffeomorphism. In this concocted example, the range of the uniform density \(\left[ 0, 1\right]\) is known beforehand. In the analogy with the cars on a highway, we would need to know the length of the highway so that our function is fully defined. Of course, we could also transport it to a uniform density on \(\left[ 0, 1\right]\) and talk about relative positions on the street.

\subsection{Limit on the number of time-delayed observations}
\label{subsec:more-delays}
\begin{figure}[htbp]
	\centering
	\includegraphics{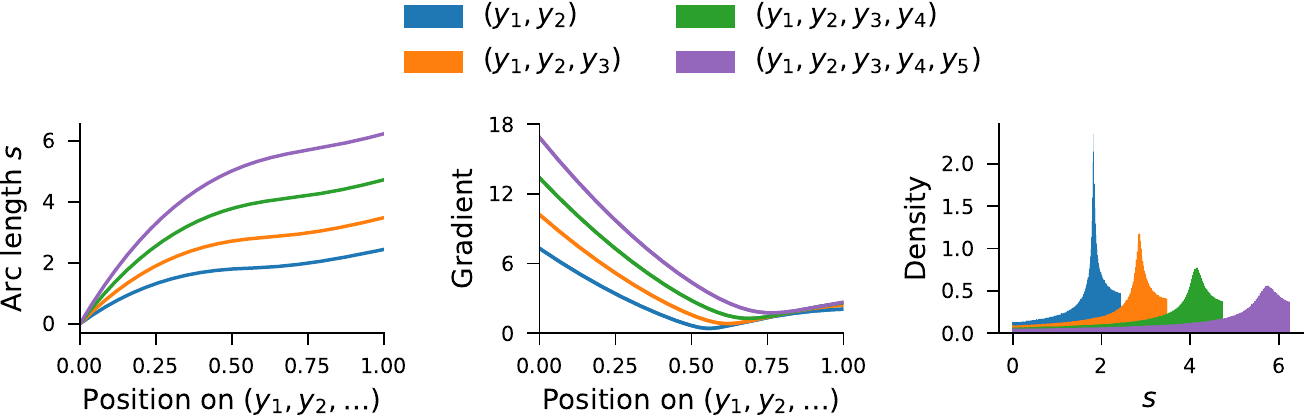}
	\caption{Effect of more time-delayed measurements.  By increasing the number of time-delayed observations, the density and the gradients get smoother. In this concrete example, an embedding can already be found with two subsequent observations (blue), while Takens' theorem limits it to three (orange). Even after that point, the properties of the densities continue to improve. }
	\label{fig:more-delays}
\end{figure}

Takens' theorem~\cite{Takens.1981} gives an upper bound for the number of time-delayed observations needed to find a diffeomorphic embedding. In practice, collecting as few measurements as possible might cut costs and reduces the size of the data set.  Still, in \cref{fig:more-delays}, we see how the densities of the arc length from the previous 1D example behave when increasing the length of the history beyond the requirement of Takens' theorem.
The gradients of the reparametrization used for the embedding and the resulting densities become smoother the longer the histories get.
This might be irrelevant in a theoretical framework but can drastically influence numerical stability and training time.
In our experiments, training the Normalizing Flow on a smooth and continuous density was more numerically stable and converged more quickly. This shows that even in scenarios with equal transport dimensions, where any transport map is sufficient, collecting measurement histories (time-delayed measurements)  can be beneficial towards improved training.

\subsection{Reconstruction of a Gaussian}\label{sec:gaussian reconstruction}
\begin{figure}[htbp]
    \centering
    \begin{subfigure}[t]{0.33\textwidth}
        \includegraphics{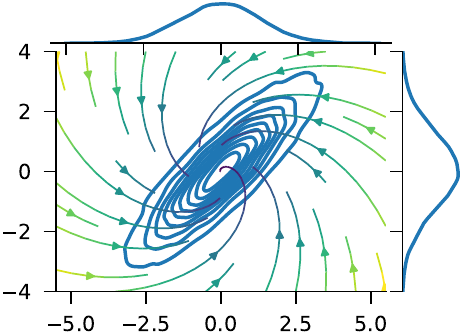}
        \caption{Joint Gaussian density}
        \label{fig:2d-gaussian-pca-example-original-density}
    \end{subfigure}
    \begin{subfigure}[t]{0.33\textwidth}
        \includegraphics{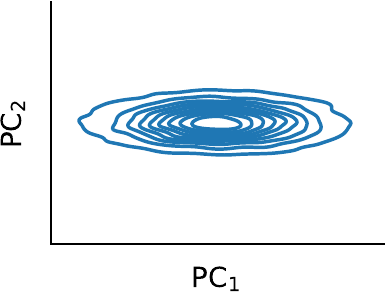}
        \caption{Reconstructed joint density}
        \label{fig:2d-gaussian-pca-example-reconstructed-density}
    \end{subfigure}
    \caption{
        Reconstruction of the normal distribution with PCA. \subref{fig:2d-gaussian-pca-example-original-density} shows a joint Gaussian density in two dimensions. The marginals are shown next to the two axes, and the streamplot shows an underlying vector field that determines the progression of sampling (collecting observations)  over time. By observing five subsequent scalar measurements of the single-dimensional $x$-coordinate of points moved over the vector field over time, the joint density can be reconstructed. \subref{fig:2d-gaussian-pca-example-reconstructed-density} depicts the reconstructed joint density in the space spanned by two principal components of the subsequent measurements of the $x$-coordinate.
    }
    \label{fig:2d-gaussian-pca-example}
\end{figure}

We now demonstrate that the concept of time-delay embedding can be used to construct a joint distribution that is in one dimension higher than the dimension of the marginal distribution the delays are sampled from. In other words, we will now demonstrate how to systematically transport from a lower-dimensional density into a higher dimension. A more detailed explanation will be in the following section.
As an example, we choose a common setup for the explanation of joint and marginal densities: a bi-variate Gaussian distribution, which is marginalized on each of the two coordinate axes. It is impossible to determine the covariance structure of the bi-variate Gaussian just from single measurements of the marginal distributions (see~\cref{fig:2d-gaussian-pca-example-original-density}). However, if we assume the coordinates are sampled from points of an underlying process (shown as streamlines over the joint distribution), we can reconstruct the full two-dimensional density (up to isometry).
Here, we record five successive $x$-coordinates of points that move over a linear vector field (streamlines in \cref{fig:2d-gaussian-pca-example-original-density}). The density of the $x$-coordinates themselves results in a normal distribution. We then compute the principal components of the new dataset, which is now embedded in a five-dimensional space. A projection to the first two principal coordinates is shown in \cref{fig:2d-gaussian-pca-example-reconstructed-density}, where the joint distribution is recovered. 
Although originally only observations in one dimension were available, the final result can now be used as the basis for an (optimal) transport in two dimensions. 

While this is a conceptual example, it is already very similar to more realistic settings. 
For example, when observing a molecule (or running molecular simulations), the molecule behaves according to underlying forces guided by an energy function. In the case where only a subset of the coordinates, or some describing variables of the molecule (e.g., bond angles), can be observed, the complete joint Boltzmann distribution can be reconstructed when sufficiently many time-delayed observations are available.

\subsection{Transporting bacteria moving on an implant  surface}\label{sec:bacteria on surface}
In the previous sections, we saw that time-delayed measurements of the observation process allow us to create a diffeomorphic version of the underlying manifold. In cases where the dimensions match, and a direct application of a transportation algorithm is feasible, this still adds the benefit of being able to find the true underlying joint distribution and might even improve the stability of the algorithm. In this section, we will demonstrate the transport to higher dimensions in detail and show that we can also transport the data to a different space on the manifold.

In the following numerical experiment, we simulate a sine-like movement of uniformly distributed bacteria on the top third of an implant surface embedded in three dimensions (see \cref{fig:cell-surface}, a spherical cap). In our test setup, a \enquote{microscope} is capable of taking two-dimensional measurements of the bacteria positions. Later, in \cref{subsec:low-to-high}, we demonstrate that one-dimensional measurements are also sufficient to reconstruct the surface.

\subsubsection{Transporting two-dimensional marginals}
\begin{figure}[htbp]
	\centering

    \begin{subfigure}[t]{0.3\textwidth}
        \includegraphics{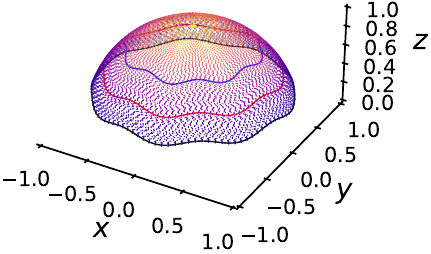}
        \caption{Bacteria on surface}
        \label{fig:cell-surface}
    \end{subfigure}
        \begin{subfigure}[t]{0.3\textwidth}
        \includegraphics{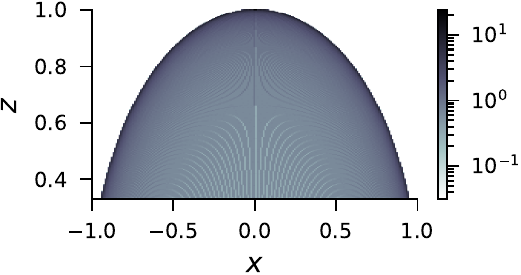}
        \caption{Histogram of observations}
        \label{fig:cell-histogram}
    \end{subfigure}

	\caption{The initial setting of the simulation. Bacteria are uniformly distributed and move on the top third of a spherical cap. In \subref{fig:cell-surface}, this is illustrated with three sample trajectories. The microscope is only capable of collecting 2D information along the $x-z$ projection. \subref{fig:cell-histogram} illustrates the resulting density of the observed $x-z$ coordinates. At the top of the sphere, the highest density is recorded.}
    \label{fig:cell}
\end{figure}

First, we assume two coordinates of the bacteria are available, namely \(x\) and \(z\).
The surface is assumed to be translucent such that we can see bacteria and record their coordinates even when they are ``on the other side" (but with the coordinate \(y\) missing). The simulated process is illustrated in \cref{fig:cell}, which also shows a histogram of the values that were recorded.

Higher-dimensional manifolds cannot be parameterized just by the arc length anymore. Adapting this step will be the major difference to the one-dimensional scenario.
We will rely on a different way to find a lower-dimensional representation of the diffeomorphic manifold embedded in a higher-dimensional space. But before, as a first step, time-delayed observations are again collected. Although five subsequent observations are sufficient, we collect six, \(( x^{(i)}_1, z^{(i)}_1, x^{(i)}_2, z^{(i)}_2, x^{(i)}_3, z^{(i)}_3 ) \), because as discussed in \cref{subsec:more-delays} more observations result in a smoother density.

As the points are now embedded in six dimensions, they cannot be analyzed visually anymore. To overcome this problem, we perform a PCA~\cite{Hotelling.1933} of the histories embedded through time delays. PCA is a linear dimensionality reduction technique that rotates the data such that the axes containing the most variance is the first principal component and so on. Higher-order principal components can then be discarded while still retaining most of the information.
 
\begin{figure}[htbp]
	\centering
    \begin{subfigure}[t]{0.3\textwidth}
	\centering
        \includegraphics{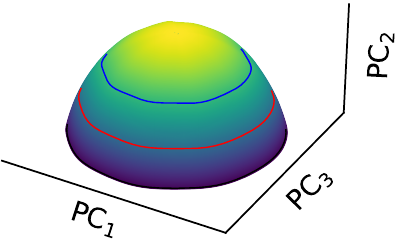}
        \caption{PCA of time-delayed observations}
        \label{fig:cell-reconstructed-pca}
    \end{subfigure}
    \begin{subfigure}[t]{0.3\textwidth}
	\centering
        \includegraphics{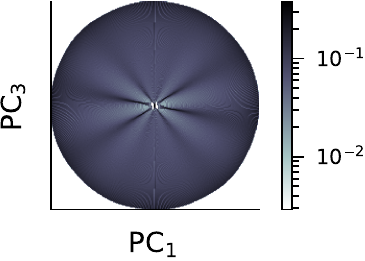}
        \caption{Transported 2D density}
        \label{fig:cell-reconstructed-density}
    \end{subfigure}
    \begin{subfigure}[t]{0.3\textwidth}
	\centering
        \includegraphics{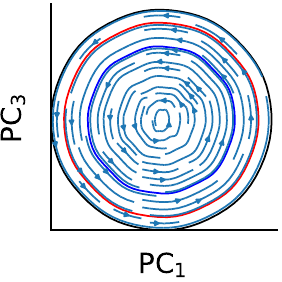}
        \caption{Bacteria movement}
        \label{fig:cell-reconstructed-streamplot}
    \end{subfigure}
    
 	\caption{Reconstruction of the implant surface with PCA. In \subref{fig:cell-reconstructed-pca}, the first three principal components of the time-delayed observations already reveal the surface's spherical shape. By visually inspecting the plot, a lower-dimensional representation can be found by looking at the surface from \enquote{above} (i.e., \(\text{PC}_1\) and \(\text{PC}_3\)). \subref{fig:cell-reconstructed-density} shows the resulting density when using those two principal components for the transport. This enables the recreation of the previously unattainable top view. In \subref{fig:cell-reconstructed-streamplot}, the movement of the bacteria has been integrated to recover the original motion.}
    \label{fig:cell-reconstructed}
    \end{figure}

The first three principal components thus contain the most variance of the data and are visualized in~\cref{fig:cell-reconstructed-pca}. The surface can already be embedded well in three dimensions by a \emph{linear} dimensionality reduction method (here, PCA of the six-dimensional vector). Thus, this simple change of taking time-delayed measurements already allowed us to transport the points into a diffeomorphic (here, almost isometric) version of the original spherical cap. By projecting the data to the first and third principal components, the top view can be reconstructed (compare~\cref{fig:cell-reconstructed-density}), even though only $x-z$ measurements were recorded originally. As we have collected a time series of the movement of the bacteria, we can use this to visualize the movement by interpolating between the points in the embedding space, as can be seen in~\cref{fig:cell-reconstructed-streamplot}. This example illustrates the benefit of time-delayed observations and that they allow for a different view of the problem---the complete shape of a diffeomorphic manifold.

\subsubsection{Transporting one-dimensional marginals}	
\label{subsec:low-to-high}
While we have already transported the two-dimensional coordinates of the bacteria into the three-dimensional space spanned by the first three Principal Components of the time-delayed measurements, the presented density transports were supported in two dimensions.
The goal for this section is to extend the previous example to the case where we transport a lower-dimensional (here, one-dimensional marginal) density to a higher-dimensional (joint) density. As the procedure relies on a diffeomorphic manifold, we can be certain that a transport to the target dimension is possible. From now on, we will adjust the setting so that we only have access to a single coordinate, \(z\), of the bacteria movement. Similarly, we collect the same number of subsequent observations \(( z^{(i)}_1, z^{(i)}_2, z^{(i)}_3, z^{(i)}_4, z^{(i)}_5, z^{(i)}_6 ) \). 
 
It is crucial to note that, in this case, the assumptions necessary for Takens' theorem do not hold anymore, because the problem is fully symmetric. With a mere up-and-down movement of the bacteria, it cannot be detected where the current bacterium is on the surface, even with time-delayed measurements.
Such perfectly symmetric scenarios are nearly ruled out in the real world since measurement equipment or the process itself has some noise or minor irregularities. 
We can overcome this problem by slightly rotating the surface randomly before applying the procedure. With this, the history of the bacteria's positions again encodes the exact position on the spherical cap. 

\begin{figure}[ht!]
	\centering
    \begin{subfigure}[t]{0.4\textwidth}
    \centering
        \includegraphics{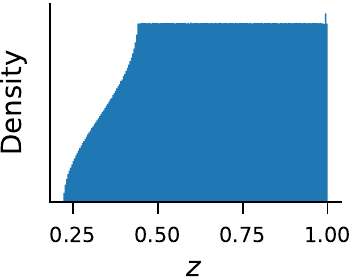}
        \caption{Histogram of observations}
        \label{fig:cell-low-to-high-density}
    \end{subfigure}
    \begin{subfigure}[t]{0.4\textwidth}
    \centering
        \includegraphics{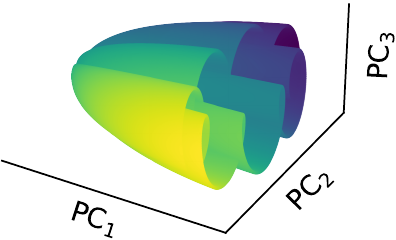}
        \caption{PCA of time-delayed observations}
        \label{fig:cell-low-to-high-pca}
    \end{subfigure}
    \begin{subfigure}[t]{0.4\textwidth}
    \centering
        \includegraphics{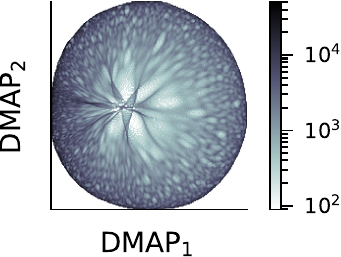}
        \caption{Transported 2D density (DMAP)}
            \label{fig:cell-low-to-high-density2d}
    \end{subfigure}
    \begin{subfigure}[t]{0.4\textwidth}
    \centering
        \includegraphics{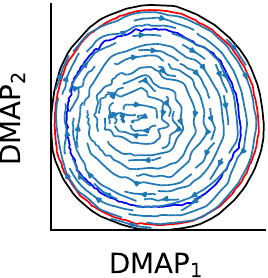}
        \caption{Reconstruced bacteria movement}
        \label{fig:cell-low-to-high-streamplot}
    \end{subfigure}
     
    \caption{Reconstruction of the implant surface with Diffusion Maps. In \subref{fig:cell-low-to-high-density}, the recorded density of $z$ is visualized. On a complete sphere, it would be uniform, but as the cap has been rotated, some values are seen less often. \subref{fig:cell-low-to-high-pca} shows the first three principal components of the time-delayed observations where no embedding is possible as the object exhibits self-intersections. 
    \subref{fig:cell-low-to-high-density2d} visualizes the reconstructed density when taking the first two DMAP vectors. It can be seen that the surface has been rotated. 
    In \subref{fig:cell-low-to-high-streamplot}, a stream plot can be reconstructed from the movement, recovering the sine-like movement of the bacteria on the surface.}
    \label{fig:cell-low-to-high}
\end{figure}

In most cases, a linear dimensionality reduction technique will not be able to recover the true underlying manifold. Above, we were able to rely on PCA to perform the density transport, which will now fail. 
The observed density for this slightly altered example is illustrated in \cref{fig:cell-low-to-high-density} with the first three principal components of the time-delayed recordings shown in \cref{fig:cell-low-to-high-pca}. As can be seen, the first three principal components are not sufficient to find an embedding for the diffeomorphic manifold, as the surface overlaps with itself.

In most cases, we need to rely on more sophisticated manifold learning techniques to embed the manifold. To achieve this, there are many different algorithms \cite{Hotelling.1933, Coifman.2006, vanderMaaten.2008, Roweis.2000} with easily attainable implementations available. In this paper, we will use Diffusion Maps (DMAPs)~\cite{Coifman.2006}. The inner workings of those algorithms are not relevant for this work, but only that those can find a lower-dimensional representation for manifolds embedded in higher dimensions.
 
Applying DMAP to the normalized PCA (to improve the numerical stability) yields comparable results to before. Again, we can use this view to complete the density transport and to reconstruct the movement of the bacteria, see \cref{fig:cell-low-to-high-density2d} and \cref{fig:cell-low-to-high-streamplot} respectively. While the top view could be reconstructed from a single variable, since we relied on a non-linear manifold learning algorithm, the result contains some numerical artifacts. Still, even though the given measurements are only from the \(z\)-coordinate, a diffeomorphic version of the two-dimensional surface could be reconstructed and the integrated movement of bacteria is continuous; cf.~\cref{fig:cell-low-to-high} \subref{fig:cell-low-to-high-density2d}, \subref{fig:cell-low-to-high-streamplot}. 

This example illustrated that time-delayed observations allow us to perform a transport from a lower to a higher dimension. Once the dimension matches the intrinsic dimension (i.e., by applying a manifold learning technique), any other density transport algorithm, such as Normalizing Flows, can be used to perform a transport in the higher dimension of the joint density.
This can be useful if we knew that the density of the joint distribution is uniform, for example. In that case, we can use classical density transport on our diffeomorphic reconstruction of the joint space to map the reconstructed joint density to another, uniform one, on our reconstructed space.

\subsection{Fusing marginals with jointly smooth functions}
\label{subsec:jointly-smooth-functions}

Up to now, we conducted experiments with a single sensor. %
In many scenarios, multiple different sensors are available to record data simultaneously. These sensors usually record the same underlying procedure, but they either measure completely different variables, observe distinct aspects, or add sensor-specific noise, which makes fusing them into one coherent joint space difficult.
Algorithms such as Canonical Correlation Analysis (CCA)~\cite{Hotelling.1936}  and their derivatives can be used to find linear dependencies between multiple observed variables.  
In the nonlinear case, one can construct jointly smooth functions (JSFs) introduced in~\cite{Dietrich.2022}, where the authors describe a procedure to process multimodal data based on manifold learning. For this, they assume that the collected data can be 
alternatively described as lying simultaneously on multiple different manifolds and determine functions that are jointly smooth on all observed manifolds. It can be shown that those jointly smooth functions represent the common features between the observations. In turn, those functions can then be used to filter out sensor-specific noise or to combine multiple sources for a more robust representation.
Imagine, for example, that multiple cameras can record the bacteria movement of before. Each camera adds sensor-specific noise and records only part of the underlying truth (i.e., only a lower-dimensional density), they all record from a different angle, and some sensors might even record additional features such as another bacteria strain. In the end, they all share the common underlying parameter space---the sphere---but finding it might be difficult.
While jointly smooth functions can find a common representation, they cannot directly determine the underlying space if its observations are non-invertible. We will use the ideas presented in this paper to overcome this issue and reconstruct the common (i.e., shared) parameter by using time-delayed measurements. For the concrete implementation of the jointly smooth functions, we used datafold~\cite{Lehmberg.2020}. Refer to appendix~\ref{apx:jsf} for the algorithm behind jointly smooth functions, and a detailed explanation of how our approach can be combined with JSFs. 

\begin{figure}[htbp]
	\centering
    \begin{subfigure}[t]{0.3\textwidth}
        \includegraphics{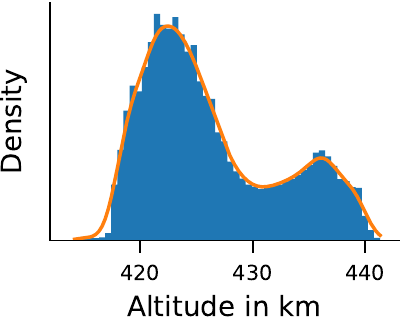}
        \caption{Altitude of ISS}
        \label{fig:satellite-altitudes-iss}
    \end{subfigure}
    \begin{subfigure}[t]{0.3\textwidth}
        \includegraphics{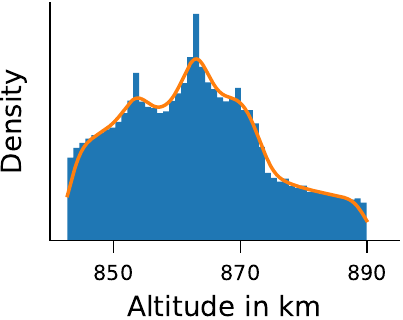}
        \caption{Altitude of NOAA-19}
        \label{fig:satellite-altitudes-noaa19}
    \end{subfigure}
    
	\caption{The observed densities in orange of the altitudes of two orbiting satellites, the ISS in \subref{fig:satellite-altitudes-iss} and NOAA-19 in \subref{fig:satellite-altitudes-noaa19}. In blue, the recovered marginal density is illustrated as a histogram.}
    \label{fig:satellite-altitudes}
\end{figure}

\begin{figure}[htbp]
	\centering
    \begin{subfigure}[t]{0.3\textwidth}
        \includegraphics{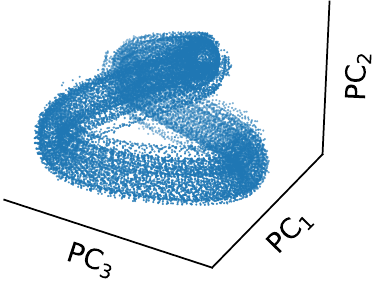}
        \caption{PCA of time-delayed observations of ISS}
        \label{fig:jsf embedding of satellite data iss}
    \end{subfigure}
    \begin{subfigure}[t]{0.3\textwidth}
        \includegraphics{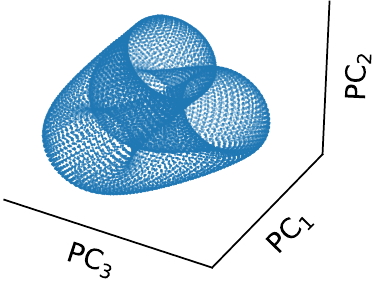}
        \caption{PCA of time-delayed observations of NOAA-19}
        \label{fig:jsf embedding of satellite data noaa19}
    \end{subfigure}
    \begin{subfigure}[t]{0.3\textwidth}
        \includegraphics{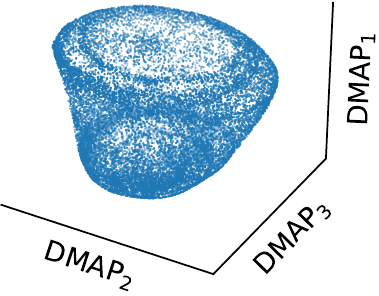}
        \caption{\label{fig:jsf satellites}Reconstructed joint distribution (DMAP)}
    \end{subfigure}
    
	\caption{The time-delayed embeddings of the ISS and NOAA-19 altitude measurements. \subref{fig:jsf embedding of satellite data iss}, and \subref{fig:jsf embedding of satellite data noaa19} show the first three principal components of 7 time-delayed altitude measurements. \subref{fig:jsf satellites} illustrates the recovered joint distribution, using Diffusion Maps on jointly smooth functions of the time-delayed embeddings of the altitude sensor data from the two satellites.}
    \label{fig:jsf embedding of satellite data}
\end{figure}

We now demonstrate that our approach can reconstruct the underlying joint distribution when observing several marginals. In this experiment, we utilize real-world scalar altitude measurements of two separate satellites, the ISS and NOAA19 (data from \url{https://celestrak.org}). Measurements of the satellites cover most of the Earth's surface. We do not utilize any of the satellites' positions but instead construct a joint embedding space of time-delayed altitude measurements.
As shown in \cref{fig:satellite-altitudes}, the marginal densities of the measured altitudes differ significantly. It is also not obvious to what degree the Earth's surface is blanketed by these measurements. For example, satellites are known to typically not cover the poles, and some of them are even geostationary and thus only cover a small constant area.
Using the jointly smooth function algorithm on the time-delayed altitude measurements allows us to reconstruct a joint embedding space, as shown in \cref{fig:jsf embedding of satellite data}.
With this joint space, it is very simple to reconstruct both original altitude measurements (depicted as histograms in~\cref{fig:satellite-altitudes}) through marginalization: For every point in~\cref{fig:jsf satellites}, we know which time-series of each altitude sensor it originated from. Hence, if we choose the first measurement in each series and construct the histogram using all points in the joint space, we obtain the marginalized altitude densities.

\section{Discussion} \label{sec:conclusion}
We discussed an approach to transport probability densities across base spaces of different dimensions. The idea is based on theoretical work~\cite{Moosmuller.2020} and extends it to the reconstruction of joint distributions from the time series of multiple marginals, but observed simultaneously. We demonstrated that solving classical optimal transport problems can fail to determine the true underlying joint distribution, especially when the map is non-invertible. By collecting time-delayed measurements of the data collection process, the joint distribution can be reconstructed, which allows for a better understanding of the underlying marginalization process. We demonstrated this approach with a simulated experiment where bacteria move on a spherical surface and a transport from 1D to 2D reconstructed the movement on the manifold.
In our approach, multiple sensor measurements can be treated as a collection of marginal distributions, and time-delayed observations together with a construction of jointly smooth features which then results in the joint distribution of all sensors.
We demonstrate this by constructing a joint distribution of altitude measurements of two satellites, namely, the ISS and NOAA19. Using this joint distribution, we can then easily reconstruct both original distributions as marginals.

Our approach can be thought of as a regularization technique for inverse problems as well. Given a process 
for dynamically sampling parameter space, and observations of the sequence of outputs from a model given the parameters, we can pinpoint exactly which point in parameter space we came from, instead of just recording indiscriminately an entire level set of possible parameters that all map to the same output.


\section*{Acknowledgements}
F.D. was funded by the German Research Foundation---project 468830823. Y.G.K. was partially supported by the US Department of Energy.
\printbibliography{}

@article{Campadelli.2015,
 author = {Campadelli, P. and Casiraghi, E. and Ceruti, C. and Rozza, A.},
 year = {2015},
 title = {Intrinsic Dimension Estimation: Relevant Techniques and a Benchmark Framework},
 pages = {1--21},
 pagination = {page},
 volume = {2015},
 issn = {1024-123X},
 journal = {Mathematical Problems in Engineering},
 doi = {10.1155/2015/759567}
}

@article{Chiappori.2017,
 author = {Chiappori, Pierre-Andr{\'e} and McCann, Robert J. and Pass, Brendan},
 year = {2017},
 title = {Multi-to One-Dimensional Optimal Transport},
 pages = {2405--2444},
 pagination = {page},
 volume = {70},
 number = {12},
 issn = {00103640},
 journal = {Communications on Pure and Applied Mathematics},
 doi = {10.1002/cpa.21707}
}

@article{Coifman.2006,
 author = {Coifman, Ronald R. and Lafon, St{\'e}phane},
 year = {2006},
 title = {Diffusion maps},
 pages = {5--30},
 pagination = {page},
 volume = {21},
 number = {1},
 issn = {10635203},
 journal = {Applied and Computational Harmonic Analysis},
 doi = {10.1016/j.acha.2006.04.006}
}

@article{Dietrich.2022,
 author = {Dietrich, Felix and Yair, Or and Mulayoff, Rotem and Talmon, Ronen and Kevrekidis, Ioannis G.},
 year = {2022},
 title = {Spectral Discovery of Jointly Smooth Features for Multimodal Data},
 pages = {410--430},
 pagination = {page},
 volume = {4},
 number = {1},
 journal = {SIAM Journal on Mathematics of Data Science},
 doi = {10.1137/21M141590X}
}

@inproceedings{Durkan.2019b,
 author = {Durkan, Conor and Bekasov, Artur and Murray, Iain and Papamakarios, George},
 title = {Neural Spline Flows},
 url = {https://proceedings.neurips.cc/paper/2019/file/7ac71d433f282034e088473244df8c02-Paper.pdf},
 volume = {32},
 publisher = {{Curran Associates, Inc}},
 editor = {Wallach, H. and Larochelle, H. and Beygelzimer, A. and Alch{\'e}-Buc, F. d$\backslash$textquotesingle and Fox, E. and Garnett, R.},
 booktitle = {Advances in Neural Information Processing Systems},
 year = {2019}
}

@article{Gangbo.2000,
 author = {Gangbo, Wilfrid and McCann, Robert J.},
 year = {2000},
 title = {Shape recognition via Wasserstein distance},
 pages = {705--737},
 pagination = {page},
 volume = {58},
 number = {4},
 issn = {0033-569X},
 journal = {Quarterly of Applied Mathematics},
 doi = {10.1090/qam/1788425}
}

@article{Hotelling.1933,
 author = {Hotelling, H.},
 year = {1933},
 title = {Analysis of a complex of statistical variables into principal components},
 pages = {417--441},
 pagination = {page},
 volume = {24},
 number = {6},
 issn = {0022-0663},
 journal = {Journal of Educational Psychology},
 doi = {10.1037/h0071325}
}

@article{Hotelling.1936,
 author = {Hotelling, Harold},
 year = {1936},
 title = {Relations Between Two Sets of Variates},
 pages = {321},
 pagination = {page},
 volume = {28},
 number = {3/4},
 issn = {00063444},
 journal = {Biometrika},
 doi = {10.2307/2333955}
}

@article{Huizing.2022,
 author = {Huizing, Geert-Jan and Peyr{\'e}, Gabriel and Cantini, Laura},
 year = {2022},
 title = {Optimal Transport improves cell-cell similarity inference in single-cell omics data},
 journal = {Bioinformatics (Oxford, England)},
 doi = {10.1093/bioinformatics/btac084}
}

@inproceedings{Kingma.2018,
 author = {Kingma, Durk P. and Prafulla, Dhariwal},
 title = {Glow: Generative Flow with Invertible 1x1 Convolutions},
 url = {https://proceedings.neurips.cc/paper/2018/file/d139db6a236200b21cc7f752979132d0-Paper.pdf},
 volume = {31},
 publisher = {{Curran Associates, Inc}},
 editor = {Bengio, S. and Wallach, H. and Larochelle, H. and Grauman, K. and Cesa-Bianchi, N. and Garnett, R.},
 booktitle = {Advances in Neural Information Processing Systems},
 year = {2018}
}

@article{Kobyzev.2021,
 author = {Kobyzev, Ivan and Prince, Simon J. D. and Brubaker, Marcus A.},
 year = {2021},
 title = {Normalizing Flows: An Introduction and Review of Current Methods},
 pages = {3964--3979},
 pagination = {page},
 volume = {43},
 number = {11},
 journal = {IEEE transactions on pattern analysis and machine intelligence},
 doi = {10.1109/TPAMI.2020.2992934}
}

@article{Lott.2016,
 author = {Lott, John},
 year = {2016},
 title = {On tangent cones in Wasserstein space},
 pages = {3127--3136},
 pagination = {page},
 volume = {145},
 number = {7},
 issn = {0002-9939},
 journal = {Proceedings of the American Mathematical Society},
 doi = {10.1090/proc/13415}
}

@article{May.2000,
 author = {May, Anthony D. and Shepherd, Simon P. and Timms, Paul M.},
 year = {2000},
 title = {Optimal transport strategies for European cities},
 pages = {285--315},
 pagination = {page},
 volume = {27},
 number = {3},
 issn = {00494488},
 journal = {Transportation},
 doi = {10.1023/A:1005274015858}
}

@article{McCann.2020,
 author = {McCann, Robert J. and Pass, Brendan},
 year = {2020},
 title = {Optimal Transportation Between Unequal Dimensions},
 pages = {1475--1520},
 pagination = {page},
 volume = {238},
 number = {3},
 issn = {0003-9527},
 journal = {Archive for Rational Mechanics and Analysis},
 doi = {10.1007/s00205-020-01569-5}
}

@article{Metivier.2016,
 author = {M{\'e}tivier, L. and Brossier, R. and M{\'e}rigot, Q. and Oudet, E. and Virieux, J.},
 year = {2016},
 title = {An optimal transport approach for seismic tomography: application to 3D full waveform inversion},
 pages = {115008},
 pagination = {page},
 volume = {32},
 number = {11},
 issn = {0266-5611},
 journal = {Inverse Problems},
 doi = {10.1088/0266-5611/32/11/115008}
}

@article{Monge.1781,
 author = {Monge, Gaspard},
 year = {1781},
 title = {M{\'e}moire sur la th{\'e}orie des d{\'e}blais et des remblais},
 pages = {666--704},
 pagination = {page},
 journal = {Histoire de l'Acad{\'e}mie Royale des Sciences de Paris}
}

@article{Moosmuller.2020,
 author = {Moosm{\"u}ller, Caroline and Dietrich, Felix and Kevrekidis, Ioannis G.},
 year = {2020},
 title = {A Geometric Approach to the Transport of Discontinuous Densities},
 pages = {1012--1035},
 pagination = {page},
 volume = {8},
 number = {3},
 journal = {SIAM/ASA Journal on Uncertainty Quantification},
 doi = {10.1137/19M1275760}
}

@article{Moriel.2021,
 author = {Moriel, Noa and Senel, Enes and Friedman, Nir and Rajewsky, Nikolaus and Karaiskos, Nikos and Nitzan, Mor},
 year = {2021},
 title = {NovoSpaRc: flexible spatial reconstruction of single-cell gene expression with optimal transport},
 pages = {4177--4200},
 pagination = {page},
 volume = {16},
 number = {9},
 journal = {Nature protocols},
 doi = {10.1038/s41596-021-00573-7}
}

@article{Philippis.2014,
 author = {de Philippis, Guido and Figalli, Alessio},
 year = {2014},
 title = {The Monge--Amp{\`e}re equation and its link to optimal transportation},
 pages = {527--580},
 pagination = {page},
 volume = {51},
 number = {4},
 issn = {0273-0979},
 journal = {Bulletin of the American Mathematical Society},
 doi = {10.1090/S0273-0979-2014-01459-4}
}

@article{Roweis.2000,
 author = {Roweis, S. T. and Saul, L. K.},
 year = {2000},
 title = {Nonlinear dimensionality reduction by locally linear embedding},
 pages = {2323--2326},
 pagination = {page},
 volume = {290},
 number = {5500},
 issn = {0036-8075},
 journal = {Science (New York, N.Y.)},
 doi = {10.1126/science.290.5500.2323}
}

@incollection{Rozza.2011,
 author = {Rozza, Alessandro and Lombardi, Gabriele and Rosa, Marco and Casiraghi, Elena and Campadelli, Paola},
 title = {IDEA: Intrinsic Dimension Estimation Algorithm},
 pages = {433--442},
 bookpagination = {page},
 volume = {6978},
 publisher = {{Springer Berlin Heidelberg}},
 isbn = {978-3-642-24084-3},
 series = {Lecture Notes in Computer Science},
 editor = {Maino, Giuseppe and Foresti, Gian Luca},
 booktitle = {Image Analysis and Processing -- ICIAP 2011},
 year = {2011},
 address = {Berlin, Heidelberg},
 doi = {10.1007/978-3-642-24085-0_45}
}

@article{Schoenberg.1946,
 author = {Schoenberg, I. J.},
 year = {1946},
 title = {Contributions to the problem of approximation of equidistant data by analytic functions. Part A. On the problem of smoothing or graduation. A first class of analytic approximation formulae},
 pages = {45--99},
 pagination = {page},
 volume = {4},
 number = {1},
 issn = {0033-569X},
 journal = {Quarterly of Applied Mathematics},
 doi = {10.1090/qam/15914}
}

@article{Schoenberg.1946b,
 author = {Schoenberg, I. J.},
 year = {1946},
 title = {Contributions to the problem of approximation of equidistant data by analytic functions. Part B. On the problem of osculatory interpolation. A second class of analytic approximation formulae},
 pages = {112--141},
 pagination = {page},
 volume = {4},
 number = {2},
 issn = {0033-569X},
 journal = {Quarterly of Applied Mathematics},
 doi = {10.1090/qam/16705}
}

@article{Tabak.2010,
 author = {Tabak, Esteban G. and Vanden-Eijnden, Eric},
 year = {2010},
 title = {Density estimation by dual ascent of the log-likelihood},
 pages = {217--233},
 pagination = {page},
 volume = {8},
 number = {1},
 issn = {15396746},
 journal = {Communications in Mathematical Sciences},
 doi = {10.4310/CMS.2010.v8.n1.a11}
}

@article{Tabak.2013,
 author = {Tabak, E. G. and Turner, Cristina V.},
 year = {2013},
 title = {A Family of Nonparametric Density Estimation Algorithms},
 pages = {145--164},
 pagination = {page},
 volume = {66},
 number = {2},
 issn = {00103640},
 journal = {Communications on Pure and Applied Mathematics},
 doi = {10.1002/cpa.21423}
}

@inproceedings{Takens.1981,
 author = {Takens, Floris},
 title = {Detecting strange attractors in turbulence},
 pages = {366--381},
 bookpagination = {page},
 volume = {898},
 publisher = {Springer},
 isbn = {978-3-540-38945-3},
 series = {Lecture Notes in Mathematics},
 editor = {Rand, David A. and Young, Lai-Sang},
 booktitle = {Dynamical Systems and Turbulence, Warwick 1980},
 year = {1981},
 address = {Berlin},
 doi = {10.1007/bfb0091924}
}

@article{vanderMaaten.2008,
 author = {{van der Maaten}, Laurens and Hinton, Geoffrey},
 year = {2008},
 title = {Visualizing Data using t-SNE},
 url = {http://jmlr.org/papers/v9/vandermaaten08a.html},
 pages = {2579--2605},
 pagination = {page},
 volume = {9},
 number = {86},
 journal = {Journal of Machine Learning Research}
}

@book{Villani.2003,
 author = {Villani, C{\'e}dric},
 year = {2003},
 title = {Topics in Optimal Transportation},
 address = {Providence, R.I. and Great Britain},
 volume = {v. 58},
 publisher = {{American Mathematical Society}},
 isbn = {0-8218-3312-X},
 series = {Graduate studies in mathematics,   1065-7339},
 institution = {{American Mathematical Society}}
}

@book{Villani.2009,
 author = {Villani, C{\'e}dric},
 year = {2009},
 title = {Optimal transport: Old and new},
 address = {Berlin and London},
 volume = {338},
 publisher = {Springer},
 isbn = {978-3-540-71050-9},
 series = {Grundlehren der mathematischen Wissenschaften,   0072-7830}
}

@article{Lehmberg.2020,
	doi       = {10.21105/joss.02283},
	url       = {https://doi.org/10.21105/joss.02283},
	year      = {2020},
	publisher = {The Open Journal},
	volume    = {5},
	number    = {51},
	pages     = {2283},
	author    = {Daniel Lehmberg and Felix Dietrich and Gerta K{\"o}ster and Hans-Joachim Bungartz},
	title     = {datafold: data-driven models for point clouds and time series on manifolds},
	journal   = {Journal of Open Source Software}}

@article{Peyre.2019,
	author = {Peyr{\'e}, Gabriel and Cuturi, Marco},
	year = {2019},
	title = {Computational Optimal Transport},
	pages = {355--607},
	pagination = {page},
	volume = {11},
	number = {5-6},
	journal = {Foundations and Trends in Machine Learning}
}

@ARTICLE{Lu2022,
author={Lu, Y. and Maulik, R. and Gao, T. and Dietrich, F. and Kevrekidis, I.G. and Duan, J.},
title={Learning the temporal evolution of multivariate densities via normalizing flows},
journal={Chaos},
year={2022},
volume={32},
number={3},
doi={10.1063/5.0065093},
art_number={033121},
note={cited By 1},
url={https://www.scopus.com/inward/record.uri?eid=2-s2.0-85126775586&doi=10.1063/5.0065093&partnerID=40&md5=bc32e21c9e8a79718cb1a1f15e423956},
document_type={Article},
source={Scopus},
}

@ARTICLE{Zheng2022,
author={Zheng, D. and Zhang, X. and Ma, K. and Bao, C.},
title={Learn From Unpaired Data for Image Restoration: A Variational Bayes Approach},
journal={IEEE Transactions on Pattern Analysis and Machine Intelligence},
year={2022},
pages={1-15},
doi={10.1109/TPAMI.2022.3215571},
note={cited By 2},
url={https://www.scopus.com/inward/record.uri?eid=2-s2.0-85140757935&doi=10.1109/TPAMI.2022.3215571&partnerID=40&md5=2dcee6cb7b0eb1303616934233d56652},
document_type={Article},
source={Scopus},
}

@ARTICLE{Bai2021,
author={Bai, Z. and Wei, H. and Xiao, Y. and Song, S. and Kucherenko, S.},
title={A vine copula-based global sensitivity analysis method for structures with multidimensional dependent variables},
journal={Mathematics},
year={2021},
volume={9},
number={19},
doi={10.3390/math9192489},
art_number={2489},
note={cited By 1},
url={https://www.scopus.com/inward/record.uri?eid=2-s2.0-85116676084&doi=10.3390/math9192489&partnerID=40&md5=fe567247ff989a4fbd5400bd623b9675},
document_type={Article},
source={Scopus},
}

@ARTICLE{Mu20201,
author={Mu, H.-Q. and Liu, H.-T. and Shen, J.-H.},
title={Copula-based uncertainty quantification (Copula-uq) for multi-sensor data in structural health monitoring},
journal={Sensors (Switzerland)},
year={2020},
volume={20},
number={19},
pages={1-18},
doi={10.3390/s20195692},
art_number={5692},
note={cited By 3},
url={https://www.scopus.com/inward/record.uri?eid=2-s2.0-85092220587&doi=10.3390/s20195692&partnerID=40&md5=8e96ca2da67d3f52ad8de850c1a0ee10},
document_type={Article},
source={Scopus},
}

@ARTICLE{Cencic2015196,
author={Cencic, O. and Frühwirth, R.},
title={A general framework for data reconciliation-Part I: Linear constraints},
journal={Computers and Chemical Engineering},
year={2015},
volume={75},
pages={196-208},
doi={10.1016/j.compchemeng.2014.12.004},
note={cited By 19},
url={https://www.scopus.com/inward/record.uri?eid=2-s2.0-85027957577&doi=10.1016/j.compchemeng.2014.12.004&partnerID=40&md5=9ba27f4679c9bf6942d7838012592c5d},
document_type={Article},
source={Scopus},
}

@CONFERENCE{Shan2011140,
author={Shan, X. and Zhou, J. and Xiao, F.},
title={Support vector machine method for multivariate density estimation based on copulas},
journal={Proceedings - 2011 International Conference on Internet Computing and Information Services, ICICIS 2011},
year={2011},
pages={140-143},
doi={10.1109/ICICIS.2011.41},
art_number={6063213},
note={cited By 3},
url={https://www.scopus.com/inward/record.uri?eid=2-s2.0-81355164045&doi=10.1109/ICICIS.2011.41&partnerID=40&md5=99fb59ff3d142080e1e75255ee24bc12},
document_type={Conference Paper},
source={Scopus},
}

@CONFERENCE{Choi20081171,
author={Choi, K.K. and Noh, Y. and Du, L.},
title={Reliability based design optimization with correlated input variables using copulas},
journal={2007 Proceedings of the ASME International Design Engineering Technical Conferences and Computers and Information in Engineering Conference, DETC2007},
year={2008},
volume={6 PART B},
pages={1171-1182},
doi={10.1115/DETC2007-35104},
note={cited By 4},
url={https://www.scopus.com/inward/record.uri?eid=2-s2.0-44949134989&doi=10.1115/DETC2007-35104&partnerID=40&md5=df93237804c814f92b73abdb974a4989},
document_type={Conference Paper},
source={Scopus},
}

@ARTICLE{Chen2005226,
author={Chen, J.-T.},
title={Using the sum-of-uniforms method to generate correlated random variates with certain marginal distribution},
journal={European Journal of Operational Research},
year={2005},
volume={167},
number={1},
pages={226-242},
doi={10.1016/j.ejor.2003.12.027},
note={cited By 12},
url={https://www.scopus.com/inward/record.uri?eid=2-s2.0-13544273196&doi=10.1016/j.ejor.2003.12.027&partnerID=40&md5=c057f95fda044e485147d59ebf954d95},
document_type={Article},
source={Scopus},
}

@ARTICLE{Oh1999411,
author={Oh, M.-S.},
title={Estimation of posterior density functions from a posterior sample},
journal={Computational Statistics and Data Analysis},
year={1999},
volume={29},
number={4},
pages={411-427},
doi={10.1016/S0167-9473(98)00068-1},
note={cited By 20},
url={https://www.scopus.com/inward/record.uri?eid=2-s2.0-0033076768&doi=10.1016/S0167-9473\%2898\%2900068-1&partnerID=40&md5=7791d159654c177ee63e335e540b9921},
document_type={Article},
source={Scopus},
}

@ARTICLE{Hogan1997239,
author={Hogan, J.W. and Laird, N.M.},
title={Mixture models for the joint distribution of repeated measures and event times},
journal={Statistics in Medicine},
year={1997},
volume={16},
number={1-3},
pages={239-257},
doi={10.1002/(sici)1097-0258(19970215)16:3<239::aid-sim483>3.0.co;2-x},
note={cited By 268},
url={https://www.scopus.com/inward/record.uri?eid=2-s2.0-0031032359&doi=10.1002/\%28sici\%291097-0258\%2819970215\%2916\%3a3\%3c239\%3a\%3aaid-sim483\%3e3.0.co\%3b2-x&partnerID=40&md5=317c4332c11aa6da13546ff6731991bf},
document_type={Conference Paper},
source={Scopus},
}

@ARTICLE{CastellanJr.1970439,
author={Castellan Jr., N.J.},
title={Determination of joint distributions from marginal distributions in dichotomous systems},
journal={Psychometrika},
year={1970},
volume={35},
number={4},
pages={439-454},
doi={10.1007/BF02291819},
note={cited By 12},
url={https://www.scopus.com/inward/record.uri?eid=2-s2.0-3042949429&doi=10.1007/BF02291819&partnerID=40&md5=00a18e1aa2e6918763a598134aa2da46},
document_type={Article},
source={Scopus},
}

@article{Whitney.1936,
  doi = {10.2307/1968482},
  url = {https://doi.org/10.2307/1968482},
  year = {1936},
  month = jul,
  publisher = {{JSTOR}},
  volume = {37},
  number = {3},
  pages = {645},
  author = {Hassler Whitney},
  title = {Differentiable Manifolds},
  journal = {The Annals of Mathematics}
}

@inproceedings{Altschuler.2017,
 author = {Altschuler, Jason and Niles-Weed, Jonathan and Rigollet, Philippe},
 booktitle = {Advances in Neural Information Processing Systems},
 editor = {I. Guyon and U. Von Luxburg and S. Bengio and H. Wallach and R. Fergus and S. Vishwanathan and R. Garnett},
 pages = {},
 publisher = {Curran Associates, Inc.},
 title = {Near-linear time approximation algorithms for optimal transport via Sinkhorn iteration},
 url = {https://proceedings.neurips.cc/paper_files/paper/2017/file/491442df5f88c6aa018e86dac21d3606-Paper.pdf},
 volume = {30},
 year = {2017}
}

@incollection{Mrigot.2021,
  doi = {10.1016/bs.hna.2020.10.001},
  url = {https://doi.org/10.1016/bs.hna.2020.10.001},
  year = {2021},
  publisher = {Elsevier},
  pages = {133--212},
  author = {Quentin M{\'{e}}rigot and Boris Thibert},
  title = {Optimal transport: discretization and algorithms},
  booktitle = {Geometric Partial Differential Equations - Part {II}}
}

@article{Kantorovich.1942,
 author = {Kantorovich, Leonid Vitaliyevich},
 year = {1942},
 title = {On the translocation of masses},
 pages = {227--229},
 pagination = {page},
 volume = {37},
 number = {7-8},
 journal = {Dokl. Acad. Nauk SSSR}
}

@article{Kantorovich.1948,
 author = {Kantorovich, Leonid Vitaliyevich},
 year = {1948},
 title = {On a Problem of Monge},
 pages = {225--226},
 pagination = {page},
 volume = {3},
 number = {2},
 journal = {Uspekhi Mat. Nauk}
}

@incollection{Courty.2014,
  doi = {10.1007/978-3-662-44848-9_18},
  url = {https://doi.org/10.1007/978-3-662-44848-9_18},
  year = {2014},
  publisher = {Springer Berlin Heidelberg},
  pages = {274--289},
  author = {Nicolas Courty and R{\'{e}}mi Flamary and Devis Tuia},
  title = {Domain Adaptation with Regularized Optimal Transport},
  booktitle = {Machine Learning and Knowledge Discovery in Databases}
}

\newpage
\appendix

\section{Normalizing Flows and Neural Spline Flows}
\label{apx:normalizing-flows}

The parametrization of transport maps is a challenge for many machine learning techniques, especially when the map must be both differentiable and invertible. A \emph{Normalizing Flow}~\cite{Tabak.2010, Tabak.2013} is a technique to define such a transport map \(T: \mathbb{R}^D \to \mathbb{R}^D\), where the respective base spaces of the transported distributions must have the same dimensionality \(D\). The key idea for the construction of a Normalizing Flow is that one of the two distributions is known (i.e., it must be efficient to sample the distribution and to evaluate its PDF). Usually, one uses a standard normal distribution, hence the name \enquote{normalizing}. With the normal distribution fixed, a Normalizing Flow maps a source distribution to the standard normal~\cite{Kobyzev.2021}. A typical construction of a Normalizing Flow approximates the transport map by a composition of a series of invertible functions~\cite{Kobyzev.2021}. 
These flows have been used in a variety of applications, e.g., to generate artificial images of faces~\cite{Kingma.2018}.

Our method does not necessarily depend on the properties of Normalizing Flows: Once the joint density has been reconstructed, any transport approach can be applied. The main advantage Normalizing Flows offer, and the reason why they have been chosen, is that they are a bijection and thus allow to sample new values and evaluate the density of points in both spaces. Although they are inherently invertible, we demonstrate that even restricted density transport algorithms can be made flexible (i.e., finding solutions of an originally non-invertible space). The specific Normalizing Flows used in this paper are so-called neural spline flows~\cite{Durkan.2019b}. They work by approximating spline knots~\cite{Schoenberg.1946, Schoenberg.1946b} of the target density with a neural network.

\section{Jointly smooth functions on non-invertible spaces}\label{apx:jsf}

\emph{Jointly Smooth Functions} (JSFs)~\cite{Dietrich.2022} are a recently developed, kernel-based data-driven approach to extract common directions in data sets. The approach constructs functions of the individual sensor data sets that are \emph{jointly smooth} across \emph{all} the available data sets. All the common functions can then be expressed in terms of these JSFs, rather than describing the common parts of each data set as functions of each of the others.

\begin{figure}[htbp]
	\centering
    \begin{subfigure}[t]{0.3\textwidth}
        \includegraphics{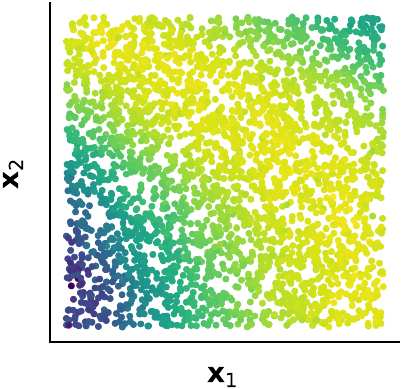}
        \caption{Sensor \(\mathbf{x}\)}
        \label{fig:jsf-base-x}
    \end{subfigure}
    \begin{subfigure}[t]{0.3\textwidth}
        \includegraphics{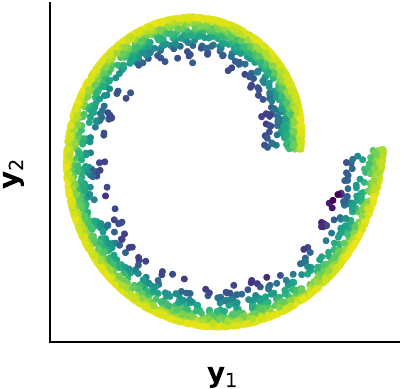}
        \caption{Sensor \(\mathbf{y}\)}
        \label{fig:jsf-base-y}
    \end{subfigure}
    
	\caption{The observations from two sensors \(\mathbf{x}\) and \(\mathbf{y}\) are shown in \subref{fig:jsf-base-x} and \subref{fig:jsf-base-y} respectively. The color is the result of the first jointly smooth function illustrating the common variable \(T\left( s \right)\). Although it is shared by both, a simpler (i.e., linear) common shared space would better describe the relation.}
	\label{fig:jsf-base}
\end{figure}

Algorithm~\ref{alg:jsf} constructs JSFs between $K$ data sets, arising from different observations of the same phenomenon, including sensor-specific (uncommon) noise. The key idea of the approximation procedure is to define function spaces on all $K$ data sets separately through eigenfunctions of kernels. Then, we use singular value decomposition (SVD) to find the ``common'' functions across these spaces. For details on this approach, see the paper by Dietrich et al.~\cite{Dietrich.2022}. In the example shown in \cref{fig:jsf-base}, we have two data sets. Therefore, we have to perform two eigendecompositions for two kernel matrices and a subsequent SVD. The ``common'' functions between the two sensors correspond to the common parameter, shown in color.

\begin{algorithm}[ht] 
    \hrulefill
    
    \textbf{\underline{Input}:} $K$ sets $\big\{ \vect{S}_{i}^{(1)},\vect{S}_{i}^{(2)},\dots\vect{S}_{i}^{(K)}\big\} _{i=1}^{N}$
    where $\vect{S}_{i}^{(k)}\in\mathbb{R}^{d_{k}}$.
    
    \textbf{\underline{Output}:} $M$ jointly smooth functions $\{ \vect{f}_{m}\in\mathbb{R}^{N}\} _{m=1}^{M}$.
    \begin{enumerate}
        \item For each observation set $\big\{ \vect{S}_{i}^{(k)}\big\} _{i=1}^{N}$
        compute the kernel:
        \[{K}_{k}(i,j)=\exp\left(-\frac{\big\Vert \vect{S}^{(k)}_{i}-\vect{S}^{(k)}_{j}\big\Vert^{2} }{2\sigma_{k}^{2}}\right)
        \]
        \item Compute $\textbf{W}_{k}\in\mathbb{R}^{N\times d}$, the first
        $d$ eigenvectors of $\textbf{K}_{k}$.
        \item Set $\textbf{W}=:\left[\textbf{W}_{1},\textbf{W}_{2},\dots,\textbf{W}_{K}\right]\in\mathbb{R}^{N\times Kd}$
        \item Compute the SVD decomposition: $\textbf{W}=\textbf{U} \textbf{$\Sigma$}\textbf{V}^{T}$
        \item Set $\vect{f}_{m}$ to be the $m$\textsuperscript{th} column of $\textbf{U}$.
    \end{enumerate}
    \caption{\label{alg:jsf}Jointly Smooth Functions from $K$ sets of observations, from~\cite{Dietrich.2022}.}
    \hrulefill
\end{algorithm}

We now demonstrate how we can combine JSFs with time-delayed observations on a small intuitive example. With the ideas presented in this paper, it can then be easily extended to higher dimensions. It is similar to what was introduced in~\cite[Section~6.1]{Dietrich.2022}, but we employ the non-invertible hook function introduced in \cref{subsec:1d-transport}. 
In this toy example, we assume a distribution where three random variables determine the observations. We consider \(( \gamma^{(i)}, \epsilon^{(i)}, \eta^{(i)} ) \sim \mathcal{U} \left[ 0, 1 \right]^3\) i.i.d. for all \(n\) observed samples. The first sensor
\begin{equation}
	\mathbf{x}^{(i)} = f \left( \gamma^{(i)}, \epsilon^{(i)} \right) =
	\begin{bmatrix}
		\gamma^{(i)}\\[\jot]
		\epsilon^{(i)}
	\end{bmatrix} \text{,}
\end{equation}
collects two variables of the system without any noise. The second sensor, on the other hand, has only access to an underlying shared variable 
\begin{equation}
	s^{(i)}  = g \left( \gamma^{(i)}, \epsilon^{(i)} \right) = \frac 1 2 \left(\gamma^{(i)}\ + {\epsilon^{(i)}}^2 \right) \text{,}
\end{equation}
which combines the previously observed parameters. Based on this shared variable it also measures sensor-specific noise, which results in the second sensor
\begin{equation}
	\mathbf{y}^{(i)} = h \left( s^{(i)}, \eta^{(i)} \right) = 
	\begin{bmatrix}
		\frac 1 2 \left(\eta^{(i)} + \frac 1 2 T\left( s^{(i)} \right) + \frac 2 3\right) \cos{\left( 2 \pi \eta^{(i)}  \right)} \\[\jot]
		\frac 1 2 \left(\eta^{(i)} + \frac 1 2 T\left( s^{(i)} \right) + \frac 2 3\right) \sin{\left( 2 \pi \eta^{(i)}  \right)}
	\end{bmatrix} \text{,}
\end{equation}
observing a spiral in \(\mathbb{R}^2\). The variable \(\eta\) specifies the angle while the shared parameter \(s\) controls the width. Note that although both sensors measure the common variable \(s\), the second sensor \(y\) only observers \(T \left( s \right) \) mapped by the non-invertible hook function \(T\left( x \right) = -2(1-x)^3 + 1.5 (1-x) + 0.5\).

Applying \cref{alg:jsf} to find JSFs yields the results illustrated in \cref{fig:jsf-base}. Although a common shared parameter could be found, since \(T\) is non-invertible, the true underlying linear parameter \(s\) cannot be reconstructed. This can be seen in the left plot illustrating the first sensor \cref{fig:jsf-base}. There, the color increases, then reaches the peak, and decays again as specified by the hook (see \cref{fig:1d-transport-map-comparison} for reference).

In a real-world context, we are unaware of the true underlying variable, and using a transformed shared variable \(s\) can be misleading and might lead to wrong results. To overcome this problem, we use time-delayed measurements to fully parameterize the points on the manifold so that the mapping becomes invertible.

For this, we assume a dynamical process and access to time-delayed measurements. Consider a nearly identical observation procedure as before with
	\begin{equation}
	\begin{bmatrix}
		\gamma_t^{(i)} \\[\jot]
		\epsilon_t^{(i)}  \\[\jot]
		\eta_t^{(i)} 
	\end{bmatrix} =
	\begin{bmatrix}
		\gamma^{(i)} + t \cdot \tau \\[\jot]
		\epsilon^{(i)} + t \cdot \tau \\[\jot]
		\eta^{(i)} + t \cdot \tau
	\end{bmatrix}
	\quad 
	t \in \{0, 1, 2\}, \tau \in \mathbb{R}
\text{,}
\end{equation}
yielding the original observations and two subsequent observations. Similarly, the sensors will observe the shifted random variables giving rise to \(\left( \mathbf{x}_0^{(i)}, \mathbf{x}_1^{(i)}, \mathbf{x}_2^{(i)} \right) \) and \(\left( \mathbf{y}_0^{(i)}, \mathbf{y}_1^{(i)}, \mathbf{y}_2^{(i)} \right) \) which we will use as observations from now on.

\begin{figure}[htbp]
	\centering
    \begin{subfigure}[t]{0.3\textwidth}
        \includegraphics{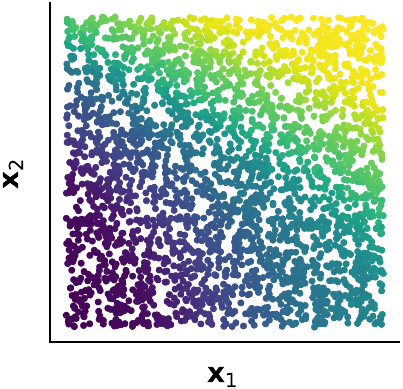}
        \caption{Sensor \(\mathbf{x}\)}
        \label{fig:jsf-reconstructed-x}
    \end{subfigure}
    \begin{subfigure}[t]{0.3\textwidth}
        \includegraphics{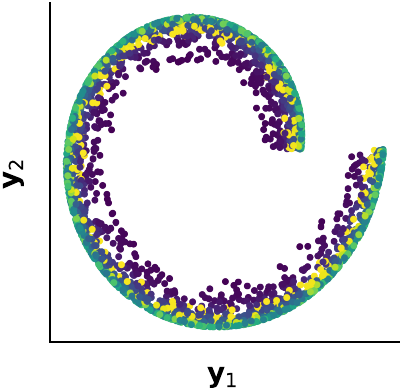}
        \caption{Sensor \(\mathbf{y}\)}
        \label{fig:jsf-reconstructed-y}
    \end{subfigure}
    
	\caption{The two sensors are depicted with the observed measurements and the color representing the shared parameter in \subref{fig:jsf-reconstructed-x} and \subref{fig:jsf-reconstructed-y} respectively. The time-delayed observations allow the function to be inverted and DMAP to find a lower-dimensional representation. With this, the true common variable \(s\) can be determined by the first DMAP axis of the first three JSFs. Plotting the shared parameter as color over sensor measurements $\mathbf{y}$ shows the non-invertibility: the points have multiple overlapping colors. A meaningful transport is only possible by using time-delayed observations.}
	\label{fig:jsf-reconstructed}
\end{figure}

The first three jointly smooth functions on this data embed the lower-dimensional manifold (the hook, 1D) into a higher-dimensional space (3D). The maximum number of subsequent observations needed is again limited by Takens' theorem. As the data is too noisy, we will not use the arc length to reparametrize the manifold but use DMAP to find a one-dimensional embedding. In the end, this gives rise to a single parameter that, up to numerical errors, is diffeomorphic to the true shared parameter \(s\) instead of \(T \left( s \right)\), see \cref{fig:jsf-reconstructed}.

With sensor $\mathbf{x}$ it can be seen that the color now correlates with the true shared parameter \(s\) as it changes linearly with the diagonal instead of a hook-like shape, as illustrated in \subref{fig:jsf-reconstructed-x}. This behaves as we would expect from \(g(\boldsymbol{\cdot},\boldsymbol{\cdot})\). Although the function \(T\) is non-linear, with our approach, we could transport the data into a higher dimension in which the manifold could be parametrized and then unfolded with DMAP into a single dimension, revealing the true shared parameter $s$.

\end{document}